\documentclass[10pt]{article}
\usepackage{graphicx}
\usepackage{amsmath}
\usepackage{amssymb}
\usepackage{caption2}
\usepackage{slashed}
\begin{document}
\bibliographystyle{prsty}
\begin{center}
{\large {\bf \sc{ The two-body strong decays of the fully-charm tetraquark states  }}} \\[2mm]
Zhi-Gang Wang$^*$\footnote{E-mail: zgwang@aliyun.com.  }, Xiao-Song Yang$^{*\dag}$    \\
 Department of Physics, North China Electric Power University, Baoding 071003, P. R. China$^*$
 School of Nuclear Science and Engineering, North China Electric Power University, Beijing 102206, P. R. China$^\dag$
\end{center}

\begin{abstract}
We study the hadronic coupling constants in the two-body strong decays of the fully-charm tetraquark states with the $J^{PC}=0^{++}$, $1^{+-}$ and $2^{++}$ via the QCD sum rules based on rigorous quark-hadron duality.
Then  we obtain the hadronic coupling constants and partial decay widths therefore total decay
widths, which support assigning the $X(6552)$ as the first radial excitation of the scalar tetraquark state. And other  predictions can be used to diagnose  exotic states  in the future.
\end{abstract}



\section{Introduction}

In 2020, the LHCb collaboration explored  the di-$J/\psi$ invariant  mass spectrum  using the $pp$ collision data at centre-of-mass energies of $\sqrt{s}=7$, $8$ and $13\,\rm{TeV}$, which corresponds  to an integrated luminosity of $9\, \rm{fb}^{-1}$  \cite{LHCb-cccc-2006}. They observed a narrow structure at about 6.9 GeV and a broad structure  above the di-$J/\psi$ threshold ranging from 6.2 to 6.8 GeV, and they also observed  some vague structures around 7.2 GeV. It is the first time to observe the fully-heavy tetraquark candidates.

In 2023, the ATLAS collaboration explored the di-$J/\psi$ and $J/\psi\psi^\prime$ invariant  mass spectra  and observed several resonances ($R$) using the $pp$ collision   data at  centre-of-mass energy  of $\sqrt{s}=13\,\rm{TeV}$, which corresponds  to an integrated luminosity of $140\, \rm{fb}^{-1}$ \cite{ATLAS-cccc-2023}. The fitted Breit-Wigner masses and widths are
\begin{flalign}
 & R_0 : M = 6.41\pm 0.08_{-0.03}^{+0.08} \mbox{ GeV}\, , \, \Gamma = 0.59\pm 0.35_{-0.20}^{+0.12} \mbox{ GeV} \, , \nonumber\\
 & R_1 : M = 6.63\pm 0.05_{-0.01}^{+0.08} \mbox{ GeV}\, , \, \Gamma = 0.35\pm 0.11_{-0.04}^{+0.11} \mbox{ GeV} \, , \nonumber\\
 & R_2 : M = 6.86\pm 0.03_{-0.02}^{+0.01} \mbox{ GeV}\, , \, \Gamma = 0.11\pm 0.05_{-0.01}^{+0.02} \mbox{ GeV} \, ,
\end{flalign}
or
\begin{flalign}
 & R_0 : M = 6.65\pm 0.02_{-0.02}^{+0.03} \mbox{ GeV}\, , \, \Gamma = 0.44\pm 0.05_{-0.05}^{+0.06} \mbox{ GeV} \, , \nonumber\\
 & R_2 : M = 6.91\pm 0.01\pm 0.01 \mbox{ GeV}\, , \, \Gamma = 0.15\pm 0.03\pm 0.01 \mbox{ GeV} \, ,
\end{flalign}
in the di-$J/\psi$ mass spectrum, and
\begin{flalign}
 & R_3 : M = 7.22\pm 0.03_{-0.04}^{+0.01} \mbox{ GeV}\, , \, \Gamma = 0.09\pm 0.06_{-0.05}^{+0.06} \mbox{ GeV} \, ,
\end{flalign}
or
\begin{flalign}
 & R_3 : M = 6.96\pm 0.05\pm 0.03 \mbox{ GeV}\, , \, \Gamma = 0.51\pm 0.17_{-0.10}^{+0.11} \mbox{ GeV} \, ,
\end{flalign}
in the $J/\psi \psi^\prime$ mass spectrum \cite{ATLAS-cccc-2023}.

While at the ICHEP 2022 conference, the ATLAS collaboration reported evidences of several fully-charm tetraquark  excesses in the di-$J/\psi$ and $J/\psi\psi^\prime$ invariant mass spectra  using the $pp$ collision data at centre-of-mass energy of $\sqrt{s}=13\,\rm{TeV}$, which corresponds to an integrated luminosity of $139\,\rm{fb}^{-1}$ \cite{Atlas2022}. The fitted Breit-Wigner masses and widths are
\begin{flalign}
 & R_0 : M = 6.22 \pm 0.05 {}^{+0.04}_{-0.05}  \mbox{ GeV}\, , \, \Gamma = 0.31 \pm 0.12{}^{+0.07}_{-0.08}  \mbox{ GeV} \, , \nonumber\\
 & R_1 : M = 6.62 \pm 0.03{}^{+0.02}_{-0.01}  \mbox{ GeV}\, , \, \Gamma = 0.31 \pm 0.09{}^{+0.06}_{-0.11} \mbox{ GeV} \, , \nonumber\\
 & R_2 : M = 6.87 \pm 0.03{}^{+0.06}_{-0.01} \mbox{ GeV}\, , \, \Gamma = 0.12 \pm 0.04{}^{+0.03}_{-0.01}  \mbox{ GeV} \, ,
\end{flalign}
in the di-$J/\psi$ mass spectrum \cite{Atlas2022}.

Also in 2023, the CMS collaboration studied the di-$J/\psi$  mass spectrum using the $pp$ collision data centre-of-mass energies of $\sqrt{s} = 13\,\rm{TeV}$, which corresponds to an integrated luminosity of
$135\,\rm{ fb}^{-1}$ \cite{CMS-cccc-2023}.  They observed three new resonance structures,  the fitted Breit-Wigner masses and widths are
\begin{flalign}
 & R_1 : M = 6552\pm10\pm12 \mbox{ MeV}\, , \, \Gamma = 124^{+32}_{-26}\pm33 \mbox{ MeV} \, , \nonumber\\
 & R_2 : M = 6927\pm9\pm4 \mbox{ MeV}\, , \, \Gamma = 122^{+24}_{-21}\pm18 \mbox{ MeV} \, , \nonumber\\
 & R_3 : M = 7287^{+20}_{-18}\pm5 \mbox{ MeV}\, , \, \Gamma = 95^{+59}_{-40}\pm19 \mbox{ MeV} \, ,
\end{flalign}
the local significances of the three peaks are $6.5$, $9.4$ and $4.1$ standard deviations, respectively.

Up to now, only the $X(6900)$ is established, however, the quantum numbers $J^{PC}$ have not been determined  yet.
The predicted masses of the fully-heavy  tetraquark (molecular) states from the  different theoretical approaches   before and after the LHCb experiment \cite{LHCb-cccc-2006} lie either  above or below the $J/\psi J/\psi$ or $\Upsilon\Upsilon$ threshold, and vary at a large range \cite{Rosner-2017,WZG-cccc-EPJC,Chen-2017,Wu-2018,FKGuo-2018-Anwar,Polosa-2018,Hughes-2018,Navarra-2019,Bai-2016,
WZG-cccc-APPB,Roberts-2020,
WZG-cccc-CPC,PingJL-2020,Zhong-2019,WangJZ-Produ-mass,Zhong-2021,Zhu-2021-NPB,WZG-cccc-IJMPA,WZG-cccc-NPB,ZhangJR-PRD,QiaoCF-2021,
GuoFK-2021-PRL,DWC-2023-PRD,YGL-2023-EPJC,XKDong-SB-2021,Gong-Zhong-2022,X6600-Azizi,X6900-Azizi}, and none of them are fully consistent with the more precise  measurements of the ATLAS and CMS collaborations \cite{ATLAS-cccc-2023,CMS-cccc-2023}.
It is certainly that  the new ATLAS and CMS experimental data would  provide more robust  promises to decipher the novel peaking structures appeared in the  di-$J/\psi$ invariant mass spectrum, and  serve as more  efficacious examinations  of various theoretical models and would be used to diagnose the nature of  the exotic $X$, $Y$ and $Z$ states.

In Refs.\cite{WZG-cccc-EPJC,WZG-cccc-APPB}, we study the mass spectrum of the scalar, axialvector, vector and tensor fully-heavy tetraquark states in the framework of  the QCD sum rules. Subsequently, we explore the  mass spectrum of the first radial excitations of the fully-charm  tetraquark states, and   resort to the Regge trajectories to obtain  the masses of the second radial excitations, and make possible assignments of the LHCb's new resonances \cite{WZG-cccc-CPC}.  In Ref.\cite{WZG-cccc-NPB},  we re-study   the mass spectrum of the ground state and first/second/third radial excited diquark-antidiquark type fully-charm  tetraquark states, then we take account of the new CMS and ATLAS experimental data  and try to make possible assignments of the $X(6600)$, $X(6900)$ and $X(7300)$ consistently.

Because we cannot assign a hadron by the mass alone, we should explore the decays  at least.  In the present work, we explore the two-body strong decays of the ground states and first radial excitations of the
scalar, axialvector and tensor fully-charm tetraquark states in the framework of the QCD sum rules based on rigorous quark-hadron duality, which works well in studying the hadronic coupling constants \cite{WZG-ZJX-Zc-Decay,WZG-Y4660-Decay,WZG-Zcs3985-decay,WZG-Zcs4123-decay,WZG-Y4500-decay},  and make more robust assignments based on the masses and widths together.

The article is organized as follows: we obtain the QCD sum rules for the hadronic coupling constants of the fully-charm tetraquark states in Section 2; in Section 3, we present the numerical results and discussions; and Section 4 is reserved for our conclusion.

\section{ QCD sum rules for the hadronic coupling constants }

Let us write down the three-point correlation functions firstly,
\begin{eqnarray}\label{CF-0}
\Pi^1(p,q) &=& i^2 \int d^4 x d^4 y e^{ip\cdot x} e^{iq\cdot y} \langle0| T\left\{J^{\eta_c}(x) J^{\eta_c}(y) J^{0\dagger}(0)\right\}|0\rangle \, , \nonumber\\
\Pi^2_{\alpha\beta}(p,q) &=& i^2 \int d^4 x d^4 y e^{ip\cdot x} e^{iq\cdot y} \langle0| T\left\{J^{J/\psi}_{\alpha}(x) J^{J/\psi}_{\beta}(y) J^{0\dagger}(0)\right\}|0\rangle \, , \nonumber\\
\Pi^3_{\alpha}(p,q) &=& i^2 \int d^4 x d^4 y e^{ip\cdot x} e^{iq\cdot y} \langle0| T\left\{J^{\chi_c}_{\alpha}(x) J^{\eta_c}(y) J^{0\dagger}(0)\right\}|0\rangle \, ,
\end{eqnarray}
\begin{eqnarray}\label{CF-1}
\Pi^4_{\mu\alpha\beta}(p,q) &=& i^2 \int d^4 x d^4 y e^{ip\cdot x} e^{iq\cdot y} \langle0| T\left\{J^{J/\psi}_{\mu}(x) J^{\eta_c}(y) J_{\alpha\beta}^{1\dagger}(0)\right\}|0\rangle \, , \nonumber\\
\Pi^5_{\mu\nu\alpha\beta}(p,q) &=& i^2 \int d^4 x d^4 y e^{ip\cdot x} e^{iq\cdot y} \langle0|
T\left\{J^{h_c}_{\mu\nu}(x) J^{\eta_c}(y) J_{\alpha\beta}^{1\dagger}(0)\right\}|0\rangle \, ,
\end{eqnarray}
\begin{eqnarray}\label{CF-2}
\Pi^6_{\alpha\beta}(p,q) &=& i^2 \int d^4 x d^4 y e^{ip\cdot x} e^{iq\cdot y} \langle0| T\left\{J^{\eta_c}(x) J^{\eta_c}(y) J_{\alpha\beta}^{2\dagger}(0) \right\}|0\rangle \, , \nonumber\\
\Pi^7_{\mu\nu\alpha\beta}(p,q) &=& i^2 \int d^4 x d^4 y e^{ip\cdot x} e^{iq\cdot y} \langle0| T\left\{J^{J/\psi}_{\mu}(x) J^{J/\psi}_{\nu}(y) J_{\alpha\beta}^{2\dagger}(0)\right\}|0\rangle \, ,
\end{eqnarray}
where the currents
\begin{eqnarray}
J^{\eta_c}(x) &=& \bar{c}(x)i\gamma_5 c(x)\, , \nonumber\\
J^{J/\psi}_{\alpha}(x) &=& \bar{c}(x)\gamma_\alpha c(x)\, , \nonumber\\
J^{\chi_c}_{\alpha}(x) &=& \bar{c}(x)\gamma_\alpha \gamma_5 c(x)\, , \nonumber\\
J^{h_c}_{\alpha\beta}(x) &=& \bar{c}(x)\sigma_{\alpha\beta} c(x)\, ,
\end{eqnarray}
interpolate the charmonia  $\eta_c$, $J/\psi$, $\chi_{c1}$ and $h_c$, respectively,  the currents
\begin{eqnarray}
J^0(x) &=& \epsilon^{ijk} \epsilon^{imn} c^T_j(x) C \gamma_\alpha c_k(x) \bar{c}_m(x)\gamma^\alpha C \bar{c}^T_n(x)\, , \nonumber\\
J_{\alpha\beta}^1(x) &=& \epsilon^{ijk} \epsilon^{imn} \Big\{c^T_j(x) C \gamma_\alpha c_k(x) \bar{c}_m(x)\gamma_\beta C \bar{c}^T_n(x) -c^T_j(x) C \gamma_\beta c_k(x) \bar{c}_m(x)\gamma_\alpha C \bar{c}^T_n(x) \Big\}\, ,\nonumber\\
J_{\alpha\beta}^2(x) &=& \frac{\epsilon^{ijk} \epsilon^{imn}}{\sqrt{2}} \Big\{c^T_j(x) C \gamma_\alpha c_k(x) \bar{c}_m(x)\gamma_\beta C \bar{c}^T_n(x) +c^T_j(x) C \gamma_\beta c_k(x) \bar{c}_m(x)\gamma_\alpha C \bar{c}^T_n(x) \Big\}\, ,
\end{eqnarray}
interpolate the fully-charm tetraquark states with the spins $0$, $1$ and $2$, respectively \cite{WZG-cccc-APPB,WZG-cccc-CPC,WZG-cccc-NPB}, the $i$, $j$, $k$, $m$ and $n$ are color indexes. We take those correlation functions to study the hadronic coupling constants $G_{X_0^{(\prime)}\eta_c\eta_c}$,
 $G_{X_0^{(\prime)}J/\psi J/\psi}$, $G_{X_0^{(\prime)}\chi_c \eta_c}$,
$G_{X_1^{(\prime)}J/\psi \eta_c}$, $G_{X_1^{(\prime)}h_c \eta_c}$,
$G_{X_2^{(\prime)}\eta_c\eta_c}$, and $G_{X_2^{(\prime)}J/\psi J/\psi}$, respectively.

At the hadron side of the correlation functions, see Eqs.\eqref{CF-0}-\eqref{CF-2}, we insert  a complete set of intermediate hadronic states with the same quantum numbers  as the currents, and isolate the ground state contributions in the charmonium channels, and the ground state plus first radial excited state contributions in the tetraquark channels explicitly \cite{SVZ79,Reinders85},
\begin{eqnarray}\label{Hadron-CT-1}
\Pi^1(p,q)&=& \frac{\lambda_{X_0} f_{\eta_c}^2 m_{\eta_c}^4 G_{X_0\eta_c \eta_c} }{4m_c^2(m_{X_0}^2-p^{\prime2})(m_{\eta_c}^2-p^2)(m_{\eta_c}^2-q^2)}\,p\cdot q \nonumber\\
&&+\frac{\lambda_{X^\prime_0} f_{\eta_c}^2 m_{\eta_c}^4 G_{X^\prime_0\eta_c \eta_c} }{4m_c^2(m_{X^\prime_0}^2-p^{\prime2})(m_{\eta_c}^2-p^2)(m_{\eta_c}^2-q^2)}\,p\cdot q     + \cdots\, , \nonumber\\
&=&\Pi_{1}(p^{\prime2},p^2,q^2)\,  p\cdot q  + \cdots\, ,
\end{eqnarray}

\begin{eqnarray}\label{Hadron-CT-2}
\Pi^2_{\alpha\beta}(p,q)&=& \frac{\lambda_{X_0} f_{J/\psi}^2 m_{J/\psi}^2 G_{X_0J/\psi J/\psi} }{(m_{X_0}^2-p^{\prime2})(m_{J/\psi}^2-p^2)(m_{J/\psi}^2-q^2)}\,  g_{\alpha\beta} \nonumber\\
&&+\frac{\lambda_{X^\prime_0} f_{J/\psi}^2 m_{J/\psi}^2 G_{X^\prime_0J/\psi J/\psi}  }{(m_{X^\prime_0}^2-p^{\prime2})(m_{J/\psi}^2-p^2)(m_{J/\psi}^2-q^2)}\,  g_{\alpha\beta}    + \cdots\, , \nonumber\\
&=&\Pi_{2}(p^{\prime2},p^2,q^2)\,   g_{\alpha\beta}  + \cdots\, ,
\end{eqnarray}

\begin{eqnarray}\label{Hadron-CT-3}
\Pi^3_{\alpha}(p,q)&=& -\frac{\lambda_{X_0} f_{\chi_c} m_{\chi_c} f_{\eta_c}m_{\eta_c}^2 G_{X_0\chi_c \eta_c}   }{2m_c(m_{X_0}^2-p^{\prime2})(m_{\chi_c}^2-p^2)(m_{\eta_c}^2-q^2)}\,iq_\alpha\nonumber\\
&&-\frac{\lambda_{X^\prime_0} f_{\chi_c} m_{\chi_c} f_{\eta_c}m_{\eta_c}^2 G_{X^\prime_0\chi_c \eta_c}   }{2m_c(m_{X^\prime_0}^2-p^{\prime2})(m_{\chi_c}^2-p^2)(m_{\eta_c}^2-q^2)}\,iq_\alpha   + \cdots\, , \nonumber\\
&=&\Pi_{3}(p^{\prime2},p^2,q^2)\, (-iq_\alpha)    + \cdots\, ,
\end{eqnarray}

\begin{eqnarray}
P_{A}^{\alpha\beta\alpha^\prime\beta^\prime}(p^\prime)\,\epsilon_{\alpha^\prime\beta^\prime}{}^{\mu\tau}\,p_\tau\,\Pi^4_{\mu\alpha\beta}(p,q)&=&
\widetilde{\Pi}_4(p^{\prime2},p^2,q^2)\left(p^2+p\cdot q\right)\, ,
\end{eqnarray}

\begin{eqnarray}\label{Hadron-CT-4}
\Pi_4(p^{\prime2},p^2,q^2)&=&\widetilde{\Pi}_4(p^{\prime2},p^2,q^2)\,p^2\, ,\nonumber\\
&=&\frac{\lambda_{X_1} f_{J/\psi}m^3_{J/\psi}f_{\eta_c} m_{\eta_c}^2 G_{X_1J/\psi \eta_c} }{2m_cm_{X_1}(m_{X_1}^2-p^{\prime2})(m_{J/\psi}^2-p^2)(m_{\eta_c}^2-q^2)}\nonumber\\
&&+\frac{\lambda_{X_1^\prime} f_{J/\psi}m^3_{J/\psi}f_{\eta_c} m_{\eta_c}^2 G_{X^\prime_1J/\psi \eta_c} }{2m_cm_{X^\prime_1}(m_{X^\prime_1}^2-p^{\prime2})(m_{J/\psi}^2-p^2)(m_{\eta_c}^2-q^2)}   + \cdots\, ,
\end{eqnarray}

\begin{eqnarray}
P_{A}^{\mu\nu\mu^\prime\nu^\prime}(p)P_{A}^{\alpha\beta\alpha^\prime\beta^\prime}(p^\prime)\,\epsilon_{\mu^\prime\nu^\prime\alpha^\prime\beta^\prime}\,
\Pi^5_{\mu\nu\alpha\beta}(p,q)&=&i\,
\widetilde{\Pi}_5(p^{\prime2},p^2,q^2)\left(-p^2q^2+(p\cdot q)^2\right)\, ,
\end{eqnarray}

\begin{eqnarray}\label{Hadron-CT-5}
\Pi_5(p^{\prime2},p^2,q^2)&=&\widetilde{\Pi}_5(p^{\prime2},p^2,q^2)\,p^2q^2\, ,\nonumber\\
&=&\frac{\lambda_{X_1} f_{h_c}m^2_{h_c}f_{\eta_c} m_{\eta_c}^4 G_{X_1h_c \eta_c} }{9m_cm_{X_1}(m_{X_1}^2-p^{\prime2})(m_{h_c}^2-p^2)(m_{\eta_c}^2-q^2)}\nonumber\\
&&+\frac{\lambda_{X^\prime_1} f_{h_c}m^2_{h_c}f_{\eta_c} m_{\eta_c}^4 G_{X^\prime_1h_c \eta_c} }{9m_cm_{X^\prime_1}(m_{X^\prime_1}^2-p^{\prime2})(m_{h_c}^2-p^2)(m_{\eta_c}^2-q^2)}   + \cdots\, ,
\end{eqnarray}

\begin{eqnarray}\label{Hadron-CT-6}
\Pi^6_{\alpha\beta}(p,q)&=&- \frac{\lambda_{X_2} f_{\eta_c}^2 m_{\eta_c}^4 (m_{X_2}^2-m_{\eta_c}^2)\, G_{X_2\eta_c \eta_c}  }{6m_c^2m_{X_2}^2(m_{X_2}^2-p^{\prime2})(m_{\eta_c}^2-p^2)(m_{\eta_c}^2-q^2)}\,p_\alpha q_\beta \nonumber\\
&&- \frac{\lambda_{X^\prime_2} f_{\eta_c}^2 m_{\eta_c}^4 (m_{X^\prime_2}^2-m_{\eta_c}^2)\, G_{X_2^\prime\eta_c \eta_c}  }{6m_c^2m_{X^\prime_2}^2(m_{X^\prime_2}^2-p^{\prime2})(m_{\eta_c}^2-p^2)(m_{\eta_c}^2-q^2)}\,p_\alpha q_\beta    + \cdots\, , \nonumber\\
&=&\Pi_{6}(p^{\prime2},p^2,q^2)\left(-p_{\alpha}q_{\beta}\right)  + \cdots\, ,
\end{eqnarray}

\begin{eqnarray}\label{Hadron-CT-7}
\Pi^7_{\mu\nu\alpha\beta}(p,q)&=&- \frac{\lambda_{X_2} f_{J/\psi}^2 m_{J/\psi}^2 \, G_{X_2J/\psi J/\psi}  }{2(m_{X_2}^2-p^{\prime2})(m_{J/\psi}^2-p^2)(m_{J/\psi}^2-q^2)}\,
\left( g_{\mu\alpha}g_{\nu\beta}+g_{\mu\beta}g_{\nu\alpha}\right) \nonumber\\
&&- \frac{\lambda_{X^\prime_2} f_{J/\psi}^2 m_{J/\psi}^2 \, G_{X^\prime_2J/\psi J/\psi}  }{2(m_{X^\prime_2}^2-p^{\prime2})(m_{J/\psi}^2-p^2)(m_{J/\psi}^2-q^2)}\,
\left( g_{\mu\alpha}g_{\nu\beta}+g_{\mu\beta}g_{\nu\alpha}\right)    + \cdots\, , \nonumber\\
&=&\Pi_{7}(p^{\prime2},p^2,q^2)\left( -g_{\mu\alpha}g_{\nu\beta}-g_{\mu\beta}g_{\nu\alpha}\right)  + \cdots\, ,
\end{eqnarray}
where
\begin{eqnarray}
P_{A}^{\mu\nu\alpha\beta}(p)&=&\frac{1}{6}\left( g^{\mu\alpha}-\frac{p^\mu p^\alpha}{p^2}\right)\left( g^{\nu\beta}-\frac{p^\nu p^\beta}{p^2}\right)\, ,
\end{eqnarray}
the decay constants or pole residues are defined by,
\begin{eqnarray}
\langle0|J^{\eta_c}(0)|\eta_c(p)\rangle&=&\frac{f_{\eta_c} m_{\eta_c}^2}{2m_c}  \,\, , \nonumber \\
\langle0|J_{\mu}^{J/\psi}(0)|J/\psi(p)\rangle&=&f_{J/\psi} m_{J/\psi} \,\xi_\mu \,\, , \nonumber \\
\langle0|J_{\mu\nu}^{h_c}(0)|h_c(p)\rangle&=&f_{h_c} \epsilon_{\mu\nu\alpha\beta}\, p^\alpha \xi^\beta \,\, , \nonumber \\
\langle0|J_{\mu}^{\chi_c}(0)|\chi_c(p)\rangle&=&f_{\chi_c} m_{\chi_c}\, \zeta_\mu \,\, ,
\end{eqnarray}
\begin{eqnarray}
 \langle 0|J^0 (0)|X^{(\prime)}_0 (p)\rangle &=& \lambda_{X^{(\prime)}_0}     \, , \nonumber\\
  \langle 0|J^1_{\mu\nu}(0)|X^{(\prime)}_1(p)\rangle &=& \tilde{\lambda}_{X^{(\prime)}_1} \, \epsilon_{\mu\nu\alpha\beta} \, \varepsilon^{\alpha}p^{\beta}\, , \nonumber\\
   \langle 0|J^2_{\mu\nu}(0)|X^{(\prime)}_2 (p)\rangle &=& \lambda_{X^{(\prime)}_2} \, \varepsilon_{\mu\nu}   \, ,
\end{eqnarray}
$\tilde{\lambda}_{X^{(\prime)}_1}m_{X^{(\prime)}_1}=\lambda_{X^{(\prime)}_1}$,
the hadronic coupling constants are defined by,
\begin{eqnarray}
\langle \eta_c(p)\eta_c(q)|X_0^{(\prime)}(p^\prime)\rangle&=& i p\cdot q \,G_{X_0^{(\prime)}\eta_c\eta_c}\, ,\nonumber \\
\langle J/\psi(p)J/\psi(q)|X_0^{(\prime)}(p^\prime)\rangle&=& i  \xi^* \cdot \xi^* \, G_{X_0^{(\prime)}J/\psi J/\psi}\, ,\nonumber \\
\langle \chi_c(p)\eta_c(q)|X_0^{(\prime)}(p^\prime)\rangle&=& -\zeta^* \cdot q \,G_{X_0^{(\prime)}\chi_c \eta_c}\, ,
\end{eqnarray}
\begin{eqnarray}
\langle J/\psi(p)\eta_c(q)|X_1^{(\prime)}(p^\prime)\rangle&=& i\xi^* \cdot \varepsilon \,G_{X_1^{(\prime)}J/\psi \eta_c}\, ,\nonumber \\
\langle h_c(p)\eta_c(q)|X_1^{(\prime)}(p^\prime)\rangle&=& \epsilon^{\lambda\tau\rho\sigma}p_{\lambda}\xi^*_{\tau}p^\prime_\rho\varepsilon_\sigma  \,G_{X_1^{(\prime)}h_c \eta_c}\, ,
\end{eqnarray}
\begin{eqnarray}
\langle \eta_c(p)\eta_c(q)|X_2^{(\prime)}(p^\prime)\rangle&=& -i \varepsilon_{\mu\nu}p^{\mu}q^{\nu} \,G_{X_2^{(\prime)}\eta_c\eta_c}\, ,\nonumber \\
\langle J/\psi(p)J/\psi(q)|X_2^{(\prime)}(p^\prime)\rangle&=& -i \varepsilon^{\alpha\beta} \xi^*_\alpha  \xi^*_\beta \, G_{X_2^{(\prime)}J/\psi J/\psi}\, ,
\end{eqnarray}
the $\xi_\mu$, $\zeta_\mu$, $\varepsilon_{\mu}$ and $\varepsilon_{\mu\nu} $ are the  polarization vectors of the corresponding charmonium/tetraquark states.
As the currents $J_{\alpha\beta}^{h_c}(x)$ and $J^1_{\alpha\beta}(x)$ couple potentially to  the charmonia/tetraquarks  with both the $J^{PC}=1^{+-}$ and $1^{--}$, we introduce the projector $P_{A}^{\mu\nu\alpha\beta}(p)$ to project out the states with the  $J^{PC}=1^{+-}$ \cite{WZG-cccc-APPB,WZG-cccc-CPC,WZG-cccc-NPB}.
In this work, we explore the hadronic coupling constants with the components $\Pi_{i}(p^{\prime2},p^2,q^2)$ with $i=1$, $\cdots$, $7$ to avoid contaminations.

Generally speaking, if a quark current and a hadron have the same quantum numbers, then they have non-vanishing couplings with each other. For example, the currents $J^0(x)$, $J^1_{\alpha\beta}(x)$ and $J^2_{\alpha\beta}(x)$ couple potentially to the 1S, 2S, $\cdots$ states of the fully-charm tetraquarks, we take into account the 1S and 2S states in the two-point QCD sum rules, and obtain the masses and poles residues by solving an univariate quadratic equation. For the technical details, one can consult Refs.\cite{WZG-cccc-CPC,WZG-cccc-NPB}.

Now, we obtain the hadronic  spectral densities $\rho_H(s^\prime,s,u)$ through triple  dispersion relation,
\begin{eqnarray}\label{dispersion-3}
\Pi_{H}(p^{\prime2},p^2,q^2)&=&\int_{16m_c^2}^\infty ds^{\prime} \int_{4m_c^2}^\infty ds \int_{4m_c^2}^\infty du \frac{\rho_{H}(s^\prime,s,u)}{(s^\prime-p^{\prime2})(s-p^2)(u-q^2)}\, ,
\end{eqnarray}
where
\begin{eqnarray}
\rho_{H}(s^\prime,s,u)&=&{\lim_{\epsilon_3\to 0}}\,\,{\lim_{\epsilon_2\to 0}} \,\,{\lim_{\epsilon_1\to 0}}\,\,\frac{ {\rm Im}_{s^\prime}\, {\rm Im}_{s}\,{\rm Im}_{u}\,\Pi_{H}(s^\prime+i\epsilon_3,s+i\epsilon_2,u+i\epsilon_1) }{\pi^3} \, ,
\end{eqnarray}
we add the subscript $H$ to stand for  the components $\Pi_{i}(p^{\prime2},p^2,q^2)$ with $i=1-7$ at the hadron side. Although the variables $p^\prime$, $p$ and $q$ have the relation $p^\prime=p+q$, it is feasible to take the $p^{\prime2}$, $p^2$ and $q^2$ as free parameters to determine the spectral densities, and we can obtain a nonzero  imaginary part indeed for all the variables $p^{\prime2}$, $p^2$ and $q^2$.

\begin{figure}
 \centering
  \includegraphics[totalheight=5cm,width=14cm]{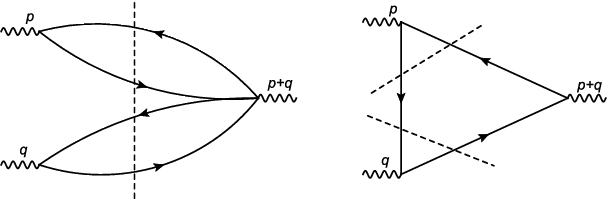}
              \caption{The lowest order Feynman diagrams  for the hadronic coupling constants $G_{X_0\eta_c \eta_c}$ and $G_{\chi_{c0}\eta_c\eta_c}$, respectively, where the dashed lines denote  the Cutkosky's cuts.} \label{fig-Feynman}
\end{figure}

At the QCD side, we take account of the perturbative terms and gluon condensate contributions, and obtain the QCD spectral densities of the components $\Pi_{i}(p^{\prime2},p^2,q^2)$ through double dispersion relation,
\begin{eqnarray}\label{dispersion-2}
\Pi_{QCD}(p^{\prime2},p^2,q^2)&=& \int_{4m_c^2}^\infty ds \int_{4m_c^2}^\infty du \frac{\rho_{QCD}(p^{\prime2},s,u)}{(s-p^2)(u-q^2)}\, ,
\end{eqnarray}
as
\begin{eqnarray}
{\rm lim}_{\epsilon \to 0}{\rm Im}\,\Pi_{QCD}(s^\prime+i\epsilon_3,p^2,q^2)&=&0\, ,
\end{eqnarray}
we cannot obtain a nonzero  imaginary part  for  the variable $p^{\prime2}$. In Fig.\ref{fig-Feynman}, we plot
the lowest order Feynman diagrams  for the hadronic coupling constants $G_{X_0\eta_c \eta_c}$ and $G_{\chi_{c0}\eta_c\eta_c}$ as an example. We calculate those diagrams with the Cutkosky's rules, although the left diagram for the tetraquark state $X_0$ does not allow a nonzero  imaginary part with respect to the $s^\prime+i\epsilon_3$, it does not forbid putting the $s^\prime$ on the mass-shell at the hadron side; while the right diagram for the traditional charmonium $\chi_{c0}$ does forbid putting the $s^\prime$ on the mass-shell at the hadron side. Thus, in the QCD sum rules for the traditional mesons, we have to put one of the variables $p^{\prime2}$, $p^2$ and $q^2$ off mass-shell.

At the hadron side, there is triple dispersion relation, see Eq.\eqref{dispersion-3}, while at the QCD side, there is only double dispersion relation, see Eq.\eqref{dispersion-2}, they do not match with each other channel by channel without performing some trick.    We accomplish   the integral over $ds^\prime$ firstly at the hadron side, then match the hadron side with the QCD side of the components $\Pi_{i}(p^{\prime2},p^2,q^2)$ of the correlation functions  bellow the continuum thresholds $s_0$ and $u_0$ to obtain rigorous quark-hadron  duality  \cite{WZG-ZJX-Zc-Decay,WZG-Y4660-Decay},
 \begin{eqnarray}\label{Duality}
  \int_{4m_c^2}^{s_0}ds \int_{4m_c^2}^{u_0}du  \left[ \int_{16m_c^2}^{\infty}ds^\prime  \frac{\rho_H(s^\prime,s,u)}{(s^\prime-p^{\prime2})(s-p^2)(u-q^2)} \right] &=&\int_{4m_c^2}^{s_{0}}ds \int_{4m_c^2}^{u_0}du  \frac{\rho_{QCD}(s,u)}{(s-p^2)(u-q^2)}
  \, .   \nonumber\\
\end{eqnarray}
Now we write down the hadron representation explicitly,
\begin{eqnarray}\label{HS-CT-1}
\Pi_{1}(p^{\prime2},p^2,q^2)&=& \frac{\lambda_{X_0} f_{\eta_c}^2 m_{\eta_c}^4 G_{X_0\eta_c \eta_c} }{4m_c^2(m_{X_0}^2-p^{\prime2})(m_{\eta_c}^2-p^2)(m_{\eta_c}^2-q^2)}\nonumber\\
&&+\frac{\lambda_{X^\prime_0} f_{\eta_c}^2 m_{\eta_c}^4 G_{X^\prime_0\eta_c \eta_c} }{4m_c^2(m_{X^\prime_0}^2-p^{\prime2})(m_{\eta_c}^2-p^2)(m_{\eta_c}^2-q^2)} +\frac{C_1 }{(m_{\eta_c}^2-p^2)(m_{\eta_c}^2-q^2)} \, ,
\end{eqnarray}

\begin{eqnarray}\label{HS-CT-2}
\Pi_{2}(p^{\prime2},p^2,q^2)&=& \frac{\lambda_{X_0} f_{J/\psi}^2 m_{J/\psi}^2 G_{X_0J/\psi J/\psi} }{(m_{X_0}^2-p^{\prime2})(m_{J/\psi}^2-p^2)(m_{J/\psi}^2-q^2)}\nonumber\\
&&  +\frac{\lambda_{X^\prime_0} f_{J/\psi}^2 m_{J/\psi}^2 G_{X^\prime_0J/\psi J/\psi}  }{(m_{X^\prime_0}^2-p^{\prime2})(m_{J/\psi}^2-p^2)(m_{J/\psi}^2-q^2)}   +\frac{C_2 }{(m_{J/\psi}^2-p^2)(m_{J/\psi}^2-q^2)} \, ,
\end{eqnarray}

\begin{eqnarray}\label{HS-CT-3}
\Pi_{3}(p^{\prime2},p^2,q^2)&=& \frac{\lambda_{X_0} f_{\chi_c} m_{\chi_c} f_{\eta_c}m_{\eta_c}^2 G_{X_0\chi_c \eta_c}   }{2m_c(m_{X_0}^2-p^{\prime2})(m_{\chi_c}^2-p^2)(m_{\eta_c}^2-q^2)}\nonumber\\
&&+\frac{\lambda_{X^\prime_0} f_{\chi_c} m_{\chi_c} f_{\eta_c}m_{\eta_c}^2 G_{X^\prime_0\chi_c \eta_c}  }{2m_c(m_{X^\prime_0}^2-p^{\prime2})(m_{\chi_c}^2-p^2)(m_{\eta_c}^2-q^2)}  + \frac{C_3}{(m_{\chi_c}^2-p^2)(m_{\eta_c}^2-q^2)}\, ,
\end{eqnarray}

\begin{eqnarray}\label{HS-CT-4}
\Pi_4(p^{\prime2},p^2,q^2)&=&\frac{\tilde{\lambda}_{X_1} f_{J/\psi}m^3_{J/\psi}f_{\eta_c} m_{\eta_c}^2 G_{X_1J/\psi \eta_c} }{2m_c(m_{X_1}^2-p^{\prime2})(m_{J/\psi}^2-p^2)(m_{\eta_c}^2-q^2)}\nonumber\\
&&+\frac{\tilde{\lambda}_{X_1^\prime} f_{J/\psi}m^3_{J/\psi}f_{\eta_c} m_{\eta_c}^2 G_{X^\prime_1J/\psi \eta_c} }{2m_c(m_{X^\prime_1}^2-p^{\prime2})(m_{J/\psi}^2-p^2)(m_{\eta_c}^2-q^2)}
+ \frac{C_4}{(m_{J/\psi}^2-p^2)(m_{\eta_c}^2-q^2)}\, ,
\end{eqnarray}

\begin{eqnarray}\label{HS-CT-5}
\Pi_5(p^{\prime2},p^2,q^2)&=&\frac{\tilde{\lambda}_{X_1} f_{h_c}m^2_{h_c}f_{\eta_c} m_{\eta_c}^4 G_{X_1h_c \eta_c} }{9m_c(m_{X_1}^2-p^{\prime2})(m_{h_c}^2-p^2)(m_{\eta_c}^2-q^2)}\nonumber\\
&&+\frac{\tilde{\lambda}_{X^\prime_1} f_{h_c}m^2_{h_c}f_{\eta_c} m_{\eta_c}^4 G_{X^\prime_1h_c \eta_c} }{9m_c(m_{X^\prime_1}^2-p^{\prime2})(m_{h_c}^2-p^2)(m_{\eta_c}^2-q^2)}   +\frac{C_5}{(m_{h_c}^2-p^2)(m_{\eta_c}^2-q^2)} \, ,
\end{eqnarray}

\begin{eqnarray}\label{HS-CT-6}
\Pi_{6}(p^{\prime2},p^2,q^2)&=& \frac{\lambda_{X_2} f_{\eta_c}^2 m_{\eta_c}^4 (m_{X_2}^2-m_{\eta_c}^2)\, G_{X_2\eta_c \eta_c}  }{6m_c^2m_{X_2}^2(m_{X_2}^2-p^{\prime2})(m_{\eta_c}^2-p^2)(m_{\eta_c}^2-q^2)} \nonumber\\
&&+ \frac{\lambda_{X^\prime_2} f_{\eta_c}^2 m_{\eta_c}^4 (m_{X^\prime_2}^2-m_{\eta_c}^2)\, G_{X_2^\prime\eta_c \eta_c}  }{6m_c^2m_{X^\prime_2}^2(m_{X^\prime_2}^2-p^{\prime2})(m_{\eta_c}^2-p^2)(m_{\eta_c}^2-q^2)}   +  \frac{C_6}{(m_{\eta_c}^2-p^2)(m_{\eta_c}^2-q^2)} \, ,
\end{eqnarray}

\begin{eqnarray}\label{HS-CT-7}
\Pi_{7}(p^{\prime2},p^2,q^2)&=& \frac{\lambda_{X_2} f_{J/\psi}^2 m_{J/\psi}^2 \, G_{X_2J/\psi J/\psi}  }{2(m_{X_2}^2-p^{\prime2})(m_{J/\psi}^2-p^2)(m_{J/\psi}^2-q^2)} \nonumber\\
&&+ \frac{\lambda_{X^\prime_2} f_{J/\psi}^2 m_{J/\psi}^2 \, G_{X^\prime_2J/\psi J/\psi}  }{2(m_{X^\prime_2}^2-p^{\prime2})(m_{J/\psi}^2-p^2)(m_{J/\psi}^2-q^2)}
+ \frac{C_7}{(m_{J/\psi}^2-p^2)(m_{J/\psi}^2-q^2)}\, ,
\end{eqnarray}
where we introduce the parameters $C_{i}$ with $i=1-7$ to represent  all the contributions concerning  the higher resonances  in the $s^\prime$ channel.

We set $p^{\prime2}=2p^2$ in the components $\Pi_H(p^{\prime 2},p^2,q^2)$, and  perform  double Borel transform in regard  to  $P^2=-p^2$ and $Q^2=-q^2$, respectively. The spectral densities $\rho_{H}(s^\prime,s,u)$ and $\rho_{QCD}(s,u)$ are physical, while the variables $p^{\prime2}$, $p^2$ and $q^2$ in Eq.\eqref{Duality} are free variables, as we perform the operator product expansion at the large space-like regions $-p^2 \to \infty$ and $-q^2 \to \infty$. Generally speaking, we can set $p^{\prime2}=\alpha p^2$ or $\alpha q^2$ with $\alpha$ to be a finite quantity. Considering the mass poles  at $s^\prime=m^2_{X^{(\prime)}_{0,1,2}}$, $s=m^2_{\eta_c,J/\psi,\chi_c,h_c}$ and $u=m^2_{\eta_c,J/\psi,\chi_c,h_c}$ have the relations $s^\prime=4s=4u$ approximately, we can set $\alpha=1-4$. In calculations, we observe that the optimal value $\alpha=2$.

 Then we set $T_1^2=T_2^2=T^2$  to obtain  seven  QCD sum rules,
\begin{eqnarray}\label{QCDSR-X0eta-eta}
&&\frac{\lambda_{X_0\eta_c \eta_c} G_{X_0\eta_c\eta_c}}{2 (\widetilde{m}^2_{X_0}-m^2_{\eta_c})}\left[\exp\left(-\frac{m^2_{\eta_c}}{T^2}\right) -\exp\left(-\frac{\widetilde{m}^2_{X_0}}{T^2}\right) \right] \exp\left(-\frac{m^2_{\eta_c}}{T^2}\right)+ \nonumber\\
&&\frac{\lambda_{X^\prime_0\eta_c \eta_c} G_{X^\prime_0\eta_c\eta_c}}{2 (\widetilde{m}^2_{X^\prime_0}-m^2_{\eta_c})}\left[\exp\left(-\frac{m^2_{\eta_c}}{T^2}\right) -\exp\left(-\frac{\widetilde{m}^2_{X^\prime_0}}{T^2}\right) \right] \exp\left(-\frac{m^2_{\eta_c}}{T^2}\right) +C_{1} \exp\left(-\frac{2m^2_{\eta_c}}{T^2}\right) \nonumber\\
&=&\frac{3m^2_c}{4\pi^4} \int^{s_{\eta_c}^0}_{4m^2_c} ds \int^{s_{\eta_c}^0}_{4m^2_c} du  \sqrt{1-\frac{4m^2_c}{s}} \sqrt{1-\frac{4m^2_c}{u}} \exp\left(-\frac{s+u}{T^2}\right) \nonumber\\
&&+\frac{ m^2_c }{4\pi^2}\langle\frac{\alpha_{s}GG}{\pi}\rangle \int^{s_{\eta_c}^0}_{4m^2_c} ds \int^{s_{\eta_c}^0}_{4m^2_c} du
 \frac{s  \left(s^2-10s m^2_c+20m^4_c\right) }{\sqrt{s\left(s-4m^2_c\right)}^5} \sqrt{1-\frac{4m^2_c}{u}} \exp\left(-\frac{s+u}{T^2}\right)\nonumber\\
&&+\frac{ m^2_c }{4\pi^2}\langle\frac{\alpha_{s}GG}{\pi}\rangle \int^{s_{\eta_c}^0}_{4m^2_c} ds \int^{s_{\eta_c}^0}_{4m^2_c} du \frac{u  \left(u^2-10u m^2_c+20m^4_c\right) }{\sqrt{u\left(u-4m^2_c\right)}^5} \sqrt{1-\frac{4m^2_c}{s}} \exp\left(-\frac{s+u}{T^2}\right) \nonumber\\
&&-\frac{ m^2_c }{4\pi^2}\langle\frac{\alpha_{s}GG}{\pi}\rangle \int^{s_{\eta_c}^0}_{4m^2_c} ds \int^{s_{\eta_c}^0}_{4m^2_c} du   \frac{1} {\sqrt{s\left(s-4m^2_c\right)} \sqrt{u\left(u-4m^2_c\right)}} \exp\left(-\frac{s+u}{T^2}\right) \nonumber\\
&&-\frac{ m^2_c }{4\pi^2}\langle\frac{\alpha_{s}GG}{\pi}\rangle \int^{s_{\eta_c}^0}_{4m^2_c} ds \int^{s_{\eta_c}^0}_{4m^2_c} du \frac{ \left(u-2m^2_c\right)} { \sqrt{u\left(u-4m^2_c\right)}^3} \sqrt{1-\frac{4m^2_c}{s}} \exp\left(-\frac{s+u}{T^2}\right) \nonumber\\
&&-\frac{ m^2_c }{4\pi^2}\langle\frac{\alpha_{s}GG}{\pi}\rangle \int^{s_{\eta_c}^0}_{4m^2_c} ds \int^{s_{\eta_c}^0}_{4m^2_c} du\frac{ \left(s-2m^2_c\right)} { \sqrt{s\left(s-4m^2_c\right)}^3} \sqrt{1-\frac{4m^2_c}{u}} \exp\left(-\frac{s+u}{T^2}\right) \, ,
\end{eqnarray}
where we introduce the notations,
\begin{eqnarray}
\lambda_{X^{(\prime)}_0\eta_c\eta_c}&=&=\frac{\lambda_{X^{(\prime)}_0} f^2_{\eta_c} m^4_{\eta_c}}{4m^2_c}\, ,
\end{eqnarray}
the other six QCD sum rules are given explicitly in the Appendix.
There are three unknown parameters in each QCD sum rules in Eq.\eqref{QCDSR-X0eta-eta} and Appendix,  we set the hadronic coupling constants  $G_{X_0\eta_c\eta_c}=G_{X_0^\prime \eta_c \eta_c}$, $G_{X_0J/\psi J/\psi}=G_{X_0^\prime J/\psi J/\psi}$,  $G_{X_0\chi_c\eta_c}=G_{X_0^\prime \chi_c \eta_c}$,  $G_{X_1J/\psi \eta_c}=G_{X_1^\prime J/\psi \eta_c}$, $G_{X_1h_c \eta_c}=G_{X_1^\prime h_c \eta_c}$, $G_{X_2\eta_c\eta_c}=G_{X_2^\prime \eta_c \eta_c}$, $G_{X_2J/\psi J/\psi}=G_{X_2^\prime J/\psi J/\psi}$,
and take the $C_{i}$ as free parameters and adjust the suitable values to obtain flat Borel platforms for the hadronic coupling constants \cite{WZG-ZJX-Zc-Decay,WZG-Y4660-Decay,WZG-Zcs3985-decay,WZG-Zcs4123-decay,WZG-Y4500-decay}. In calculations, we observe that there appear endpoint divergences at the thresholds  $s=4m_c^2$ and $u=4m_c^2$ due to powers of $s-4m_c^2$ and $u-4m_c^2$ in the denominators, we make the replacements $s-4m_c^2\to s-4m_c^2+\Delta^2$ and $u-4m_c^2\to u-4m_c^2+\Delta^2$ with $\Delta^2=m_c^2$ to regulate the divergences \cite{WZG-6c-IJMPA,WZG-5c-NPB}. As the gluon condensates make tiny contributions, such regulations work well.

\section{Numerical results and discussions}	
At the QCD side, we take the standard gluon condensate $\langle \frac{\alpha_s
GG}{\pi}\rangle=0.012\pm0.004\,\rm{GeV}^4$
\cite{SVZ79,Reinders85,Colangelo-Review} and  take the $\overline{MS}$ mass $m_{c}(m_c)=(1.275\pm0.025)\,\rm{GeV}$  from the Particle Data Group \cite{PDG}.
We take  account of the energy-scale dependence of  the  $\overline{MS}$ mass,
 \begin{eqnarray}
 m_c(\mu)&=&m_c(m_c)\left[\frac{\alpha_{s}(\mu)}{\alpha_{s}(m_c)}\right]^{\frac{12}{33-2n_f}} \, ,\nonumber\\
\alpha_s(\mu)&=&\frac{1}{b_0t}\left[1-\frac{b_1}{b_0^2}\frac{\log t}{t} +\frac{b_1^2(\log^2{t}-\log{t}-1)+b_0b_2}{b_0^4t^2}\right]\, ,
\end{eqnarray}
  where $t=\log \frac{\mu^2}{\Lambda^2}$, $b_0=\frac{33-2n_f}{12\pi}$, $b_1=\frac{153-19n_f}{24\pi^2}$, $b_2=\frac{2857-\frac{5033}{9}n_f+\frac{325}{27}n_f^2}{128\pi^3}$,  $\Lambda=213\,\rm{MeV}$, $296\,\rm{MeV}$  and  $339\,\rm{MeV}$ for the flavors  $n_f=5$, $4$ and $3$, respectively  \cite{PDG}.
As we study the fully-charm mesons, so we choose the flavor numbers $n_f=4$.

At the hadron side, we take $m_{\eta_c}=2.9834\,\rm{GeV}$,  $m_{J/\psi}=3.0969\,\rm{GeV}$, $m_{h_c}=3.525\,\rm{GeV}$,  $m_{\chi_{c1}}=3.51067\,\rm{GeV}$ from the Particle Data Group \cite{PDG},
$s^0_{h_c}=(3.9\,\rm{GeV})^2$, $s^0_{\chi_{c1}}=(3.9\,\rm{GeV})^2$, $s^0_{J/\psi}=(3.6\,\rm{GeV})^2$, $s^0_{\eta_c}=(3.5\,\rm{GeV})^2$,
 $f_{h_c}=0.235\,\rm{GeV}$, $f_{J/\psi}=0.418 \,\rm{GeV}$, $f_{\eta_c}=0.387 \,\rm{GeV}$   \cite{Becirevic}, $f_{\chi_{c1}}=0.338\,\rm{GeV}$ \cite{Charmonium-PRT}, $M_{X_0}=6.20\,\rm{GeV}$, $\lambda_{X_0}=2.68\times 10^{-1}\,\rm{GeV}^5$, $M_{X_1}=6.24\,\rm{GeV}$, $\lambda_{X_1}=2.18\times 10^{-1}\,\rm{GeV}^5$,      $M_{X_2}=6.27\,\rm{GeV}$, $\lambda_{X_2}=2.35\times 10^{-1}\,\rm{GeV}^5$,
  $M_{X^\prime_0}=6.57\,\rm{GeV}$,  $\lambda_{X^\prime_0}=8.12\times 10^{-1}\,\rm{GeV}^5$,
 $M_{X^\prime_1}=6.64\,\rm{GeV}$,   $\lambda_{X^\prime_1}=6.43\times 10^{-1}\,\rm{GeV}^5$,
 $M_{X^\prime_2}=6.69\,\rm{GeV}$,   $\lambda_{X^\prime_2}=6.84\times 10^{-1}\,\rm{GeV}^5$ from the QCD sum rules \cite{WZG-cccc-NPB}.  In general, we take the hadron masses, decay constants or pole residues from the two-point QCD sum rules. As it is easy to reproduce the experimental values of the charmonium masses with the QCD sum rules, we take the precise masses from the Particle Data Group for simplicity. For other hadron parameters, we take the central values, as the uncertainties originate from the input parameters at the QCD side, and the uncertainties of the continuum threshold parameters can be absorbed into the decay constants or pole residues to a large extent in the two-point QCD sum rules. Taking the central values can avoid doubly  counting the uncertainties.

\begin{figure}
 \centering
  \includegraphics[totalheight=5cm,width=7cm]{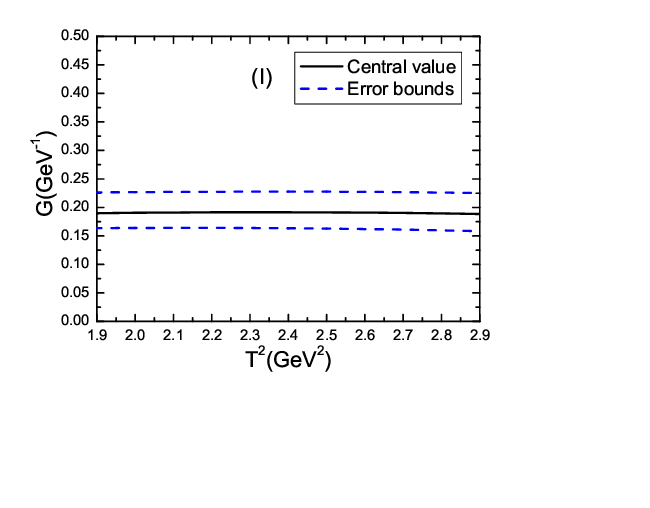}
  \includegraphics[totalheight=5cm,width=7cm]{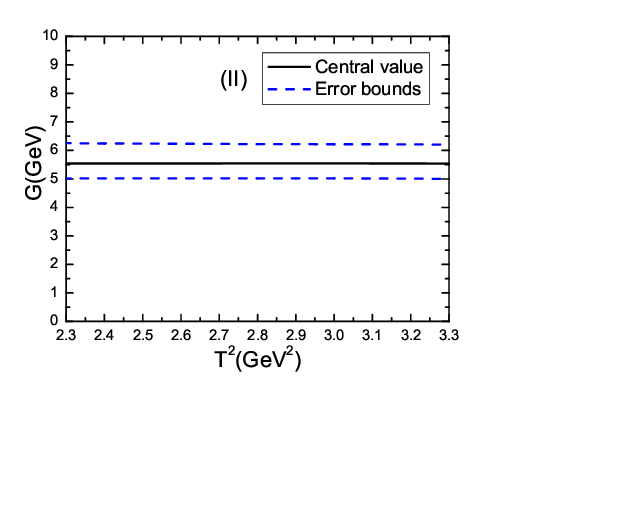}
        \caption{The central values of the hadronic coupling constants  with variations of the  Borel  parameters  $T^2$, where the (I) and (II)  denote the $G_{X_0 \eta_c \eta_c}$  and $G_{X_0 J/\psi J/\psi}$,  respectively.}\label{fig-di-eta-Jpsi}
\end{figure}

In calculations, we fit the free parameters to be
$C_{1}=0.007\,(T^2-1.0\,\rm{GeV}^2)\,\rm{GeV}^4$, $C_{2}=0.097\,(T^2-1.5\,\rm{GeV}^2)\,\rm{GeV}^6$,
$C_{3}=0.0096\,(T^2-2.0\,\rm{GeV}^2)\,\rm{GeV}^5$,
$C_4=0.29\,(T^2-1.1{\rm GeV}^2)\,{\rm GeV}^7$, $C_5=0.0017\,(T^2-2.2{\rm GeV}^2)\,{\rm GeV}^6$,
$C_6=0.0086\,(T^2-0.9{\rm GeV}^2)\,{\rm GeV}^4$, and $C_7=0.066\,(T^2-1.5{\rm GeV}^2)\,{\rm GeV}^6$,
and obtain the Borel platforms
$T^2_{X_0\eta_c\eta_c}=(1.9-2.9)\,\rm{GeV}^2$,
$T^2_{X_0J/\psi J/\psi}=(2.3-3.3)\,\rm{GeV}^2$,
$T^2_{X_0\chi_c \eta_c}=(3.1-4.1)\,\rm{GeV}^2$,
$T^2_{X_1 J/\psi \eta_c}=(2.0-3.0)\,{\rm GeV}^2$,
$T^2_{X_1 h_c \eta_c}=(2.0-3.0)\,{\rm GeV}^2$,
$T^2_{X_2 \eta_c \eta_c}=(2.4-3.4)\,{\rm GeV}^2$, and
$T^2_{X_2 J/\psi J/\psi}=(2.3-3.3)\,{\rm GeV}^2$,
where we add the subscripts $X_0\eta_c \eta_c$, $X_0J/\psi J/\psi$,
$X_0\chi_c \eta_c$, $X_1 J/\psi\eta_c$, $X_1 h_c\eta_c$, $X_2\eta_c \eta_c$ and
$X_2J/\psi J/\psi$ to denote the corresponding channels.
We obtain uniform flat platforms  $T^2_{max}-T^2_{min}=1\,\rm{GeV}^2$, where the max and min denote the maximum and minimum, respectively, just like what have been done in our previous works \cite{WZG-ZJX-Zc-Decay,WZG-Y4660-Decay,WZG-Zcs3985-decay,WZG-Zcs4123-decay,WZG-Y4500-decay}. For example, in Fig.\ref{fig-di-eta-Jpsi}, we plot the hadronic coupling constants $G_{X_0\eta_c\eta_c}$ and
 $G_{X_0J/\psi J/\psi}$  with variations of the Borel parameters at large intervals. In the Borel windows, there appear very flat platforms indeed, it is reliable to extract the hadronic  coupling constants.

Now, we estimate the uncertainties in the same way. For example, the  uncertainties of an input parameter $\xi$, $\xi= \bar{\xi} +\delta \xi$,  result in the uncertainties $\lambda_{X_0}f_{J/\psi}^2G_{X_0J/\psi J/\psi} = \bar{\lambda}_{X_0}\bar{f}_{J/\psi}^2\bar{G}_{X_0J/\psi J/\psi}
+\delta\,\lambda_{X_0}f_{J/\psi}^2G_{X_0J/\psi J/\psi}$, $C_{2} = \bar{C}_{2}+\delta C_{2}$,
\begin{eqnarray}\label{Uncertainty-4}
\delta\,\lambda_{X_0}f_{J/\psi}^2G_{X_0J/\psi J/\psi} &=&\bar{\lambda}_{X_0}\bar{f}_{J/\psi}^2\bar{G}_{X_0J/\psi J/\psi}\left( 2\frac{\delta f_{J/\psi}}{\bar{f}_{J/\psi}} +\frac{\delta \lambda_{X_0}}{\bar{\lambda}_{X_0}}+\frac{\delta G_{X_0J/\psi J/\psi}}{\bar{G}_{X_0J/\psi J/\psi}}\right)\, ,
\end{eqnarray}
we can set $\delta C_{2}=0$ and $ \frac{\delta f_{J/\psi}}{\bar{f}_{J/\psi}} = \frac{\delta \lambda_{X_0}}{\bar{\lambda}_{X_0}}=\frac{\delta G_{X_0J/\psi J/\psi}}{\bar{G}_{XJ/\psi J/\psi}}$ approximately.
Finally, we obtain the values of the hadronic coupling constants,
\begin{eqnarray} \label{HCC-values}
G_{X_0\eta_c \eta_c} &=&0.19^{+0.04}_{-0.03}\,\rm{GeV}^{-1}\, , \nonumber\\
G_{X_0J/\psi J/\psi} &=&5.54^{+0.68}_{-0.53}\,\rm{GeV}\, , \nonumber\\
G_{X_0\chi_c \eta_c} &=&0.82^{+0.09}_{-0.08}  \, , \nonumber\\
G_{X_1 J/\psi \eta_c}&=& 8.47^{+1.32}_{-1.15}\,{\rm GeV},\nonumber\\
G_{X_1 h_c \eta_c}&=& 0.30^{+0.04}_{-0.03}\,{\rm GeV}^{-1},\nonumber\\
G_{X_2 \eta_c \eta_c}&=& 0.94^{+0.17}_{-0.16}\,{\rm GeV}^{-1},\nonumber\\
G_{X_2 J/\psi J/\psi}&=& 9.69^{+1.25}_{-1.02}\,{\rm GeV}.
\end{eqnarray}

Now we set $m_{X_0}=6.20\,\rm{GeV}$, $m_{X^\prime_0}=6.57\,\rm{GeV}$,  $m_{X_1}=6.24\,\rm{GeV}$, $m_{X^\prime_1}=6.64\,\rm{GeV}$,  $m_{X_2}=6.27\,\rm{GeV}$, and $m_{X^\prime_2}=6.69\,\rm{GeV}$ based on the calculations of the QCD sum rules \cite{WZG-cccc-NPB}.
Then we obtain the partial decay widths directly,
\begin{eqnarray} \label{X0-decay}
\Gamma(X_0\to \eta_c \eta_c)&=& 3.3^{+1.6}_{-1.2} \,\rm{MeV}\, , \nonumber\\
\Gamma(X_0\to J/\psi J/\psi)&=& 13.2^{+3.5}_{-2.4} \,\rm{MeV}\, , \nonumber\\
\Gamma(X^\prime_0\to \eta_c \eta_c)&=& 7.3^{+3.5}_{-2.1} \,\rm{MeV}\, , \nonumber\\
\Gamma(X^\prime_0\to J/\psi J/\psi)&=& 110.5^{+28.7}_{-20.2} \,\rm{MeV}\, , \nonumber\\
\Gamma(X^\prime_0\to \chi_{c1} \eta_c)&=& 0.26^{+0.06}_{-0.05} \,\rm{MeV}\, ,
\end{eqnarray}

\begin{eqnarray}\label{X1-decay}
\Gamma(X_1 \rightarrow J/\psi \eta_c)&=& 52.2^{+17.5}_{-13.2}\, {\rm MeV}\, ,\nonumber\\
\Gamma(X^\prime_1 \rightarrow J/\psi \eta_c)&=& 91.7^{+30.8}_{-23.2}\,{\rm MeV}\, ,\nonumber\\
\Gamma(X^\prime_1 \rightarrow h_c \eta_c)&=& 0.67^{+0.19}_{-0.13}\,{\rm MeV}\, ,
\end{eqnarray}

\begin{eqnarray}\label{X2-decay}
\Gamma(X_2 \rightarrow \eta_c \eta_c)&=& 0.10^{+0.04}_{-0.03}\, {\rm MeV}\, ,\nonumber\\
\Gamma(X_2 \rightarrow J/\psi J/\psi)&=& 47.1^{+12.9}_{-9.4}\, {\rm MeV}\, ,\nonumber\\
\Gamma(X^\prime_2 \rightarrow \eta_c \eta_c)&=& 0.83^{+0.33}_{-0.26}\, {\rm MeV}\, ,\nonumber\\
\Gamma(X^\prime_2 \rightarrow J/\psi J/\psi)&=& 117.6^{+32.3}_{-23.5}\, {\rm MeV}\, .
\end{eqnarray}

Then we obtain the total decay  widths,
\begin{eqnarray}
\Gamma_{X_0}&=&16.6^{+6.0}_{-2.4}\,\rm{MeV}\, ,\nonumber\\
\Gamma_{X^\prime_0}&=&118.1^{+32.2}_{-22.3}\,\rm{MeV}\, ,
\end{eqnarray}
\begin{eqnarray}
\Gamma_{X_1}&=& 52.2^{+17.5}_{-13.2}\, {\rm MeV}\, ,\nonumber\\
\Gamma_{X^\prime_1}&=& 92.3^{+31.0}_{-23.3}\, {\rm MeV}\, ,
\end{eqnarray}
\begin{eqnarray}
\Gamma_{X_2}&=& 47.2^{+13.2}_{-9.4}\, {\rm MeV}\, ,\nonumber\\
\Gamma_{X^\prime_2}&=& 118.5^{+32.6}_{-23.7}\, {\rm MeV}\, .
\end{eqnarray}
by saturating  the total widths with only the supposed dominant decays, which take place through the Okubo-Zweig-Iizuka super-allowed fall-apart mechanism. In fact, such approximations would underestimate the total widths slightly, as there maybe exist other non-dominant decays, for example,
the two-body strong decays $X_0 \to D\bar{D}$ and $D^*\bar{D}^*$ can take place through
annihilating a $c\bar{c}$ pair and creating a $q\bar{q}$ pair, which are Okubo-Zweig-Iizuka forbidden.

The predicted width  $\Gamma_{X_0}=16.6^{+6.0}_{-2.4}\,\rm{MeV}$ is much smaller than the preliminary
data $\Gamma_{X(6220)}=0.31 \pm 0.12{}^{+0.07}_{-0.08}  \mbox{ GeV} $ from the ATLAS collaboration \cite{Atlas2022}. The predicted width $\Gamma_{X^\prime_0}=118.1^{+32.2}_{-22.3}\,\rm{MeV}$ is in very good agreement with the experimental data $\Gamma_{X(6552)} = 124^{+32}_{-26}\pm33 \mbox{ MeV}$ from the CMS collaboration \cite{CMS-cccc-2023}, which indicates that the $X(6552)$ can be assigned as the first radial excitation of the scalar tetraquark state. The resonances from the LHCb, ATLAS and CMS collaborations are observed in the $J/\psi J/\psi$ or $J/\psi \psi^\prime$ invariant mass spectrum, the charge conjugation should be positive, the assignments as the tetraquark states with the $J^{PC}=1^{+-}$ are not favored.

In Refs.\cite{X6600-Azizi,X6900-Azizi},   Agaev et al take the $X(6600)$ and $X(6900)$ as the ground state $A\bar{A}$ and $P\bar{P}$ type fully-charm scalar tetraquark states, respectively, and study the masses and two-body strong decays with the QCD sum rules, where the $A$ and $P$ denote the axialvector and pseudoscalar diquarks, respectively. While in Ref.\cite{WZG-cccc-NPB}, the $X(6220)$ and $X(6552)$ are assigned as the ground state and first radial excited state of the  $A\bar{A}$ type scalar tetraquark states, respectively.  By studying the two-body strong decays in the present work, we obtain further support to assign the $X(6552)$  as the first radial excitation of the scalar tetraquark state.

All the predictions  can be confronted to the experimental data in the future and serve as valuable guides for the high energy experiments.

\section{Conclusion}
In this work, we study the hadronic coupling constants in the two-body strong decays of the fully-charm tetraquark states with the $J^{PC}=0^{++}$, $1^{+-}$ and $2^{++}$ via the three-point correlation functions.
We carry out the operator product expansion up to the gluon condensate and obtain the spectral representation through double-dispersion, then match the hadron side with the QCD side to obtain rigorous quark-hadron duality.
We obtain the hadronic coupling constants and acquire the partial decay widths therefore the total decay widths
of the fully-charm tetraquark states with the $J^{PC}=0^{++}$, $1^{+-}$ and $2^{++}$, respectively. Our predictions support assigning the $X(6552)$ as the first radial excitation of the scalar tetraquark state. Other predictions play a valuable role in diagnosing the $X$, $Y$ and $Z$ states and can be confronted to the experimental data in the future.

\section*{Appendix}
The analytical expressions of the other QCD sum rules,
\begin{eqnarray}\label{QCDSR-X0psi-psi}
&&\frac{\lambda_{X_0J/\psi J/\psi} G_{X_0J/\psi J/\psi}}{2(\widetilde{m}^2_{X_0}-m^2_{J/\psi})}\left[\exp\left(-\frac{m^2_{J/\psi}}{T^2}\right) -\exp\left(-\frac{\widetilde{m}^2_{X_0}}{T^2}\right) \right] \exp\left(-\frac{m^2_{J/\psi}}{T^2}\right) +\nonumber\\
&&\frac{\lambda_{X^\prime_0J/\psi J/\psi} G_{X^\prime_0J/\psi J/\psi}}{2(\widetilde{m}^2_{X^\prime_0}-m^2_{J/\psi})}\left[\exp\left(-\frac{m^2_{J/\psi}}{T^2}\right) -\exp\left(-\frac{\widetilde{m}^2_{X^\prime_0}}{T^2}\right) \right] \exp\left(-\frac{m^2_{J/\psi}}{T^2}\right) +C_{2} \exp\left(-\frac{2m^2_{J/\psi}}{T^2}\right) \nonumber\\
&=& \frac{1}{12\pi^4}  \int^{s^0_{J/\psi}}_{4m^2_c} ds \int^{s^0_{J/\psi}}_{4m^2_c} du \left(s+2m^2_c\right) \left(u+2m^2_c\right) \sqrt{1-\frac{4m^2_c}{s}} \sqrt{1-\frac{4m^2_c}{u}} \exp\left(-\frac{s+u}{T^2}\right) \nonumber\\
&&+\frac{m^2_c } {36\pi^2}\langle\frac{\alpha_{s}GG}{\pi}\rangle \int^{s^0_{J/\psi}}_{4m^2_c} ds \int^{s^0_{J/\psi}}_{4m^2_c} du \frac{s  \left(s^2-22sm^2_c+48m^4_c\right)\left(u+2m^2_c\right)} {\sqrt{s\left(s-4m^2_c\right)}^5} \sqrt{1-\frac{4m^2_c}{u}} \exp\left(-\frac{s+u}{T^2}\right)  \nonumber\\
&&+\frac{m^2_c } {36\pi^2}\langle\frac{\alpha_{s}GG}{\pi}\rangle \int^{s^0_{J/\psi}}_{4m^2_c} ds \int^{s^0_{J/\psi}}_{4m^2_c} du \frac{u \left(u^2-22um^2_c+48m^4_c\right)\left(s+2m^2_c\right)} {\sqrt{u\left(u-4m^2_c\right)}^5} \sqrt{1-\frac{4m^2_c}{s}} \exp\left(-\frac{s+u}{T^2}\right)  \nonumber\\
&&-\frac{5m^4_c} {12\pi^2}\langle\frac{\alpha_{s}GG}{\pi}\rangle \int^{s^0_{J/\psi}}_{4m^2_c} ds \int^{s^0_{J/\psi}}_{4m^2_c} du \frac{1} {\sqrt{s\left(s-4m^2_c\right)} \sqrt{u\left(u-4m^2_c\right)}} \exp\left(-\frac{s+u}{T^2}\right)  \nonumber\\
&&-\frac{m^2_c } {36\pi^2}\langle\frac{\alpha_{s}GG}{\pi}\rangle \int^{s^0_{J/\psi}}_{4m^2_c} ds \int^{s^0_{J/\psi}}_{4m^2_c} du \frac{ \left(u-6m^2_c\right) \left(s+2m^2_c\right)} { \sqrt{u\left(u-4m^2_c\right)}^3} \sqrt{1-\frac{4m^2_c}{s}} \exp\left(-\frac{s+u}{T^2}\right)  \nonumber\\
&&-\frac{m^2_c } {36\pi^2}\langle\frac{\alpha_{s}GG}{\pi}\rangle \int^{s^0_{J/\psi}}_{4m^2_c} ds \int^{s^0_{J/\psi}}_{4m^2_c} du \frac{ \left(s-6m^2_c\right) \left(u+2m^2_c\right)} {  \sqrt{s\left(s-4m^2_c\right)}^3} \sqrt{1-\frac{4m^2_c}{u}} \exp\left(-\frac{s+u}{T^2}\right)  \, ,
\end{eqnarray}

\begin{eqnarray}\label{QCDSR-X0chi-eta}
&&\frac{\lambda_{X_0\chi_c\eta_c} G_{X_0\chi_c\eta_c}}{2 (\widetilde{m}^2_{X_0}-m^2_{\chi_c})}\left[\exp\left(-\frac{m^2_{\chi_c}}{T^2}\right) -\exp\left(-\frac{\widetilde{m}^2_{X_0}}{T^2}\right) \right] \exp\left(-\frac{m^2_{\eta_c}}{T^2}\right) +\nonumber\\
&&\frac{\lambda_{X^\prime_0\chi_c\eta_c} G_{X^\prime_0\chi_c\eta_c}}{2 (\widetilde{m}^2_{X^\prime_0}-m^2_{\chi_c})}\left[\exp\left(-\frac{m^2_{\chi_c}}{T^2}\right) -\exp\left(-\frac{\widetilde{m}^2_{X^\prime_0}}{T^2}\right) \right] \exp\left(-\frac{m^2_{\eta_c}}{T^2}\right) +C_{3} \exp\left(-\frac{m^2_{\chi_c}+m^2_{\eta_c}}{T^2}\right) \nonumber\\
&=&\frac{m_c}{4\pi^4} \int^{s^0_{\chi_c}}_{4m^2_c} ds \int^{s^0_{\eta_c}}_{4m^2_c} du \left(s-4m^2_c\right) \sqrt{1-\frac{4m^2_c}{s}} \sqrt{1-\frac{4m^2_c}{u}} \exp\left(-\frac{s+u}{T^2}\right) \nonumber\\
&&+\frac{m^3_c }{12\pi^2}\langle\frac{\alpha_{s}GG}{\pi}\rangle \int^{s^0_{\chi_c}}_{4m^2_c} ds \int^{s^0_{\eta_c}}_{4m^2_c} du \frac{\left(s+6m^2_c\right) }{\sqrt{s\left(s-4m^2_c\right)}^3} \sqrt{1-\frac{4m^2_c}{u}} \exp\left(-\frac{s+u}{T^2}\right) \nonumber\\
&&+\frac{ m_c }{12\pi^2}\langle\frac{\alpha_{s}GG}{\pi}\rangle \int^{s^0_{\chi_c}}_{4m^2_c} ds \int^{s^0_{\eta_c}}_{4m^2_c} du
\frac{u \left(s-4m^2_c\right) \left(u^2-10u m^2_c+20m^4_c\right) }{\sqrt{u\left(u-4m^2_c\right)}^5} \sqrt{1-\frac{4m^2_c}{s}} \exp\left(-\frac{s+u}{T^2}\right) \nonumber\\
&&+\frac{m^3_c} {6\pi^2} \langle\frac{\alpha_{s}GG}{\pi}\rangle \int^{s^0_{\chi_c}}_{4m^2_c} ds \int^{s^0_{\eta_c}}_{4m^2_c} du \frac{1} {\sqrt{s\left(s-4m^2_c\right)} \sqrt{u\left(u-4m^2_c\right)}} \exp\left(-\frac{s+u}{T^2}\right) \nonumber\\
&&-\frac{m_c }{12\pi^2}\langle\frac{\alpha_{s}GG}{\pi}\rangle \int^{s^0_{\chi_c}}_{4m^2_c} ds \int^{s^0_{\eta_c}}_{4m^2_c} du \frac{ \left(s-4m^2_c\right) \left(u-2m^2_c\right)} { \sqrt{u\left(u-4m^2_c\right)}^3} \sqrt{1-\frac{4m^2_c}{s}} \exp\left(-\frac{s+u}{T^2}\right) \nonumber\\
&&-\frac{m^3_c } {12\pi^2 }\langle\frac{\alpha_{s}GG}{\pi}\rangle \int^{s^0_{\chi_c}}_{4m^2_c} ds \int^{s^0_{\eta_c}}_{4m^2_c} du \frac{ \left(s-6m^2_c\right)} { \sqrt{s\left(s-4m^2_c\right)}^3} \sqrt{1-\frac{4m^2_c}{u}} \exp\left(-\frac{s+u}{T^2}\right) \, ,
\end{eqnarray}

\begin{eqnarray}\label{QCDSR-X1psi-eta}
&&\frac{\lambda_{X_1J/\psi\eta_c} G_{X_1J/\psi\eta_c}}{2 (\widetilde{m}^2_{X_1}-m^2_{J/\psi})}\left[\exp\left(-\frac{m^2_{J/\psi}}{T^2}\right) -\exp\left(-\frac{\widetilde{m}^2_{X_1}}{T^2}\right) \right] \exp\left(-\frac{m^2_{\eta_c}}{T^2}\right) +\nonumber\\
&&\frac{\lambda_{X^\prime_1J/\psi\eta_c} G_{X^\prime_1J/\psi\eta_c}}{2 (\widetilde{m}^2_{X^\prime_1}-m^2_{J/\psi})}\left[\exp\left(-\frac{m^2_{J/\psi}}{T^2}\right) -\exp\left(-\frac{\widetilde{m}^2_{X^\prime_1}}{T^2}\right) \right] \exp\left(-\frac{m^2_{\eta_c}}{T^2}\right)  +C_{4} \exp\left(-\frac{m^2_{J/\psi}+m^2_{\eta_c}}{T^2}\right) \nonumber\\
&=& \frac{ m_c }{6\pi^4}\int^{s^0_{J/\psi}}_{4m^2_c} ds \int^{s^0_{\eta_c}}_{4m^2_c} du \,u \left(2s+m^2_c\right) \sqrt{1-\frac{4m^2_c}{s}} \sqrt{1-\frac{4m^2_c}{u}} \exp\left(-\frac{s+u}{T^2}\right)\nonumber\\
&&+\frac{ m_c }{36\pi^2}\langle\frac{\alpha_{s}GG}{\pi}\rangle \int^{s^0_{J/\psi}}_{4m^2_c} ds \int^{s^0_{\eta_c}}_{4m^2_c} du \frac{s u \left(48m^6_c+38s m^4_c-29s^2 m^2_c+3s^3 \right)}{\sqrt{s\left(s-4m^2_c\right)}^5} \sqrt{1-\frac{4m^2_c}{u}} \exp\left(-\frac{s+u}{T^2}\right)\nonumber\\
&&+\frac{ m_c }{36\pi^2}\langle\frac{\alpha_{s}GG}{\pi}\rangle \int^{s^0_{J/\psi}}_{4m^2_c} ds \int^{s^0_{\eta_c}}_{4m^2_c} du \frac{u \left[8m^6_c\left(6s+5u\right)-4um^4_c\left(s+5u\right)+2u^2 m^2_c\left(u-5s\right)+su^3 \right]}{\sqrt{u\left(u-4m^2_c\right)}^5}\nonumber\\
&&\sqrt{1-\frac{4m^2_c}{s}} \exp\left(-\frac{s+u}{T^2}\right) \nonumber\\
&&+\frac{ m_c^3 }{36\pi^2}\langle\frac{\alpha_{s}GG}{\pi}\rangle \int^{s^0_{J/\psi}}_{4m^2_c} ds \int^{s^0_{\eta_c}}_{4m^2_c} du \frac{m^2_c\left(s+u\right)+s \left(s-4u\right)} {36\pi^2 s \sqrt{s\left(s-4m^2_c\right)} \sqrt{u\left(u-4m^2_c\right)}} \exp\left(-\frac{s+u}{T^2}\right) \nonumber\\
&&+\frac{ m_c }{36\pi^2}\langle\frac{\alpha_{s}GG}{\pi}\rangle \int^{s^0_{J/\psi}}_{4m^2_c} ds \int^{s^0_{\eta_c}}_{4m^2_c} du \frac{4m^4_c\left(3s+u\right)-2um^2_c\left(5s+u\right)-su^2} { \sqrt{u\left(u-4m^2_c\right)}^3} \sqrt{1-\frac{4m^2_c}{s}} \exp\left(-\frac{s+u}{T^2}\right) \nonumber\\
&&+\frac{ m_c }{36\pi^2}\langle\frac{\alpha_{s}GG}{\pi}\rangle \int^{s^0_{J/\psi}}_{4m^2_c} ds \int^{s^0_{\eta_c}}_{4m^2_c} du \frac{s^2 -3sm^2_c +6m^4_c} {\sqrt{s\left(s-4m^2_c\right)}^3} \sqrt{u\left(u-4m^2_c\right)} \exp\left(-\frac{s+u}{T^2}\right)\, ,
\end{eqnarray}

\begin{eqnarray}\label{QCDSR-X1hc-eta}
&&\frac{\lambda_{X_1h_c\eta_c} G_{X_1h_c\eta_c}}{2 (\widetilde{m}^2_{X_1}-m^2_{h_c})}\left[\exp\left(-\frac{m^2_{h_c}}{T^2}\right) -\exp\left(-\frac{\widetilde{m}^2_{X_1}}{T^2}\right) \right] \exp\left(-\frac{m^2_{\eta_c}}{T^2}\right) +\nonumber\\
&&\frac{\lambda_{X^\prime_1h_c\eta_c} G_{X^\prime_1h_c\eta_c}}{2 (\widetilde{m}^2_{X^\prime_1}-m^2_{h_c})}\left[\exp\left(-\frac{m^2_{h_c}}{T^2}\right) -\exp\left(-\frac{\widetilde{m}^2_{X^\prime_1}}{T^2}\right) \right] \exp\left(-\frac{m^2_{\eta_c}}{T^2}\right)  +C_{5} \exp\left(-\frac{m^2_{h_c}+m^2_{\eta_c}}{T^2}\right) \nonumber\\
&=& \frac{1}{144\pi^4} \int^{s^0_{h_c}}_{4m^2_c} ds \int^{s^0_{\eta_c}}_{4m^2_c} du u\left(s-4m^2_c\right) \sqrt{1-\frac{4m^2_c}{s}} \sqrt{1-\frac{4m^2_c}{u}} \exp\left(-\frac{s+u}{T^2}\right) \nonumber\\
&&+\frac{ m^2_c}{108\pi^2 }\langle\frac{\alpha_{s}GG}{\pi}\rangle \int^{s^0_{h_c}}_{4m^2_c} ds \int^{s^0_{\eta_c}}_{4m^2_c} du \frac{s u }{ \sqrt{s\left(s-4m^2_c\right)}^3} \sqrt{1-\frac{4m^2_c}{u}} \exp\left(-\frac{s+u}{T^2}\right) \nonumber\\
&&-\frac{ m^4_c}{54\pi^2 } \langle\frac{\alpha_{s}GG}{\pi}\rangle \int^{s^0_{h_c}}_{4m^2_c} ds \int^{s^0_{\eta_c}}_{4m^2_c} du \frac{u \left(s-4m^2_c\right) \left(u-2m^2_c\right)}{ \sqrt{u\left(u-4m^2_c\right)}^5} \sqrt{1-\frac{4m^2_c}{s}} \exp\left(-\frac{s+u}{T^2}\right) \nonumber\\
&&+\frac{1}{216\pi^2}\langle\frac{\alpha_{s}GG}{\pi}\rangle \int^{s^0_{h_c}}_{4m^2_c} ds \int^{s^0_{\eta_c}}_{4m^2_c} du \frac{m^4_c \left(3s-11u\right)+sm^2_c \left(u-s\right)+s^2u} { s\sqrt{s\left(s-4m^2_c\right)} \sqrt{u\left(u-4m^2_c\right)}} \exp\left(-\frac{s+u}{T^2}\right) \nonumber\\
&&-\frac{m^2_c } {108\pi^2 }\langle\frac{\alpha_{s}GG}{\pi}\rangle \int^{s^0_{h_c}}_{4m^2_c} ds \int^{s^0_{\eta_c}}_{4m^2_c} du \frac{ \left(s-4m^2_c\right) \left(u-m^2_c\right)} { \sqrt{u\left(u-4m^2_c\right)}^3} \sqrt{1-\frac{4m^2_c}{s}} \exp\left(-\frac{s+u}{T^2}\right) \nonumber\\
&&-\frac{m^4_c} {108\pi^2}\langle\frac{\alpha_{s}GG}{\pi}\rangle \int^{s^0_{h_c}}_{4m^2_c} ds \int^{s^0_{\eta_c}}_{4m^2_c} du \frac{\sqrt{u\left(u-4m^2_c\right)}} {\sqrt{s\left(s-4m^2_c\right)}^3}  \exp\left(-\frac{s+u}{T^2}\right)\, ,
\end{eqnarray}

\begin{eqnarray}\label{QCDSR-X2eta-eta}
&&\frac{\lambda_{X_2\eta_c\eta_c} G_{X_2\eta_c\eta_c}}{2 (\widetilde{m}^2_{X_2}-m^2_{\eta_c})}\left[\exp\left(-\frac{m^2_{\eta_c}}{T^2}\right) -\exp\left(-\frac{\widetilde{m}^2_{X_2}}{T^2}\right) \right] \exp\left(-\frac{m^2_{\eta_c}}{T^2}\right) +\nonumber\\
&&\frac{\lambda_{X^\prime_2\eta_c\eta_c} G_{X^\prime_2\eta_c\eta_c}}{2 (\widetilde{m}^2_{X^\prime_2}-m^2_{\eta_c})}\left[\exp\left(-\frac{m^2_{\eta_c}}{T^2}\right) -\exp\left(-\frac{\widetilde{m}^2_{X^\prime_2}}{T^2}\right) \right] \exp\left(-\frac{m^2_{\eta_c}}{T^2}\right)  +C_{6} \exp\left(-\frac{2m^2_{\eta_c}}{T^2}\right) \nonumber\\
&=&\frac{3m^2_c}{2\sqrt{2}\pi^4}\int^{s^0_{\eta_c}}_{4m^2_c} ds \int^{s^0_{\eta_c}}_{4m^2_c} du  \sqrt{1-\frac{4m^2_c}{s}} \sqrt{1-\frac{4m^2_c}{u}} \exp\left(-\frac{s+u}{T^2}\right) \nonumber\\
&&+\frac{m^2_c }{2\sqrt{2}\pi^2}\langle\frac{\alpha_{s}GG}{\pi}\rangle \int^{s^0_{\eta_c}}_{4m^2_c} ds \int^{s^0_{\eta_c}}_{4m^2_c} du \frac{s \left(s^2-10sm^2_c+20m^4_c\right)}{ \sqrt{s\left(s-4m^2_c\right)}^5} \sqrt{1-\frac{4m^2_c}{u}} \exp\left(-\frac{s+u}{T^2}\right) \nonumber\\
&&+\frac{m^2_c} {2\sqrt{2}\pi^2}\langle\frac{\alpha_{s}GG}{\pi}\rangle \int^{s^0_{\eta_c}}_{4m^2_c} ds \int^{s^0_{\eta_c}}_{4m^2_c} du \frac{u \left(u^2-10um^2_c+20m^4_c\right)} {\sqrt{u\left(u-4m^2_c\right)}^5} \sqrt{1-\frac{4m^2_c}{s}} \exp\left(-\frac{s+u}{T^2}\right) \nonumber\\
&&+\frac{m^2_c} {3\sqrt{2}\pi^2}\langle\frac{\alpha_{s}GG}{\pi}\rangle \int^{s^0_{\eta_c}}_{4m^2_c} ds \int^{s^0_{\eta_c}}_{4m^2_c} du \frac{1} {\sqrt{s\left(s-4m^2_c\right)} \sqrt{u\left(u-4m^2_c\right)}} \exp\left(-\frac{s+u}{T^2}\right) \nonumber\\
&&+\frac{m^2_c } {2\sqrt{2}\pi^2}\langle\frac{\alpha_{s}GG}{\pi}\rangle \int^{s^0_{\eta_c}}_{4m^2_c} ds \int^{s^0_{\eta_c}}_{4m^2_c} du \frac{ u-2m^2_c} { \sqrt{u\left(u-4m^2_c\right)}^3} \sqrt{1-\frac{4m^2_c}{s}} \exp\left(-\frac{s+u}{T^2}\right) \nonumber\\
&&+\frac{m^2_c } {2\sqrt{2}\pi^2}\langle\frac{\alpha_{s}GG}{\pi}\rangle \int^{s^0_{\eta_c}}_{4m^2_c} ds \int^{s^0_{\eta_c}}_{4m^2_c} du \frac{s-2m^2_c} { \sqrt{s\left(s-4m^2_c\right)}^3} \sqrt{1-\frac{4m^2_c}{u}} \exp\left(-\frac{s+u}{T^2}\right) \, ,
\end{eqnarray}

\begin{eqnarray}\label{QCDSR-X2psi-psi}
&&\frac{\lambda_{X_2J/\psi J/\psi} G_{X_2J/\psi J/\psi}}{2 (\widetilde{m}^2_{X_2}-m^2_{J/\psi})}\left[\exp\left(-\frac{m^2_{J/\psi}}{T^2}\right) -\exp\left(-\frac{\widetilde{m}^2_{X_2}}{T^2}\right) \right] \exp\left(-\frac{m^2_{J/\psi}}{T^2}\right) +\nonumber\\
&&\frac{\lambda_{X^\prime_2J/\psi J/\psi} G_{X^\prime_2J/\psi J/\psi}}{2 (\widetilde{m}^2_{X^\prime_2}-m^2_{J/\psi})}\left[\exp\left(-\frac{m^2_{J/\psi}}{T^2}\right) -\exp\left(-\frac{\widetilde{m}^2_{X^\prime_2}}{T^2}\right) \right] \exp\left(-\frac{m^2_{J/\psi}}{T^2}\right)  +C_{7} \exp\left(-\frac{2m^2_{J/\psi}}{T^2}\right) \nonumber\\
&=&\frac{1}{12\sqrt{2}\pi^4}\int^{s^0_{J/\psi}}_{4m^2_c} ds \int^{s^0_{J/\psi}}_{4m^2_c} du \left(s+2m^2_c\right) \left(u+2m^2_c\right) \sqrt{1-\frac{4m^2_c}{s}} \sqrt{1-\frac{4m^2_c}{u}} \exp\left(-\frac{s+u}{T^2}\right) \nonumber\\
&&+\frac{m^2_c} {36\sqrt{2}\pi^2}\langle\frac{\alpha_{s}GG}{\pi}\rangle \int^{s^0_{J/\psi}}_{4m^2_c} ds \int^{s^0_{J/\psi}}_{4m^2_c} du \frac{s \left(u+2m^2_c\right) \left(s^2-22sm^2_c+48m^4_c\right)} {\sqrt{s\left(s-4m^2_c\right)}^5} \sqrt{1-\frac{4m^2_c}{u}} \exp\left(-\frac{s+u}{T^2}\right) \nonumber\\
&&+\frac{m^2_c} {36\sqrt{2}\pi^2}\langle\frac{\alpha_{s}GG}{\pi}\rangle \int^{s^0_{J/\psi}}_{4m^2_c} ds \int^{s^0_{J/\psi}}_{4m^2_c} du \frac{u \left(s+2m^2_c\right) \left(u^2-22um^2_c+48m^4_c\right)} {\sqrt{u\left(u-4m^2_c\right)}^5} \sqrt{1-\frac{4m^2_c}{s}} \exp\left(-\frac{s+u}{T^2}\right) \nonumber\\
&&+\frac{m^4_c} {12\sqrt{2}\pi^2}\langle\frac{\alpha_{s}GG}{\pi}\rangle \int^{s^0_{J/\psi}}_{4m^2_c} ds \int^{s^0_{J/\psi}}_{4m^2_c} du \frac{1} {\sqrt{s\left(s-4m^2_c\right)} \sqrt{u\left(u-4m^2_c\right)}} \exp\left(-\frac{s+u}{T^2}\right) \nonumber\\
&&-\frac{m^2_c} {36\sqrt{2}\pi^2}\langle\frac{\alpha_{s}GG}{\pi}\rangle \int^{s^0_{J/\psi}}_{4m^2_c} ds \int^{s^0_{J/\psi}}_{4m^2_c} du \frac{\left(u-6m^2_c\right) \left(s+2m^2_c\right)} { \sqrt{u\left(u-4m^2_c\right)}^3} \sqrt{1-\frac{4m^2_c}{s}} \exp\left(-\frac{s+u}{T^2}\right) \nonumber\\
&&-\frac{m^2_c} {36\sqrt{2}\pi^2}\langle\frac{\alpha_{s}GG}{\pi}\rangle \int^{s^0_{J/\psi}}_{4m^2_c} ds \int^{s^0_{J/\psi}}_{4m^2_c} du \frac{ \left(s-6m^2_c\right) \left(u+2m^2_c\right)} { \sqrt{s\left(s-4m^2_c\right)}^3} \sqrt{1-\frac{4m^2_c}{u}} \exp\left(-\frac{s+u}{T^2}\right) \, ,
\end{eqnarray}
where we introduce the notations,
\begin{eqnarray}
\lambda_{X^{(\prime)}_0J/\psi J/\psi}&=&\lambda_{X^{(\prime)}_0} f^2_{J/\psi} m^2_{J/\psi}\, , \nonumber \\
\lambda_{X^{(\prime)}_0\chi_c\eta_c}&=&\frac{\lambda_{X^{(\prime)}_0} f_{\chi_c} m_{\chi_c} f_{\eta_c} m^2_{\eta_c}}{2m_c}\, ,
\end{eqnarray}
\begin{eqnarray}
\lambda_{X^{(\prime)}_1J/\psi\eta_c}&=&\frac{\lambda_{X^{(\prime)}_1} f_{J/\psi} m_{J/\psi}^3 f_{\eta_c} m^2_{\eta_c}}{2m_cm_{X^{(\prime)}_1}}\, , \nonumber\\
\lambda_{X^{(\prime)}_1h_c\eta_c}&=&\frac{\lambda_{X^{(\prime)}_1} f_{h_c} m_{h_c}^2 f_{\eta_c} m^4_{\eta_c}}{9m_cm_{X^{(\prime)}_1}}\, ,
\end{eqnarray}
\begin{eqnarray}
\lambda_{X^{(\prime)}_2\eta_c\eta_c}&=&\frac{\lambda_{X^{(\prime)}_2} f_{\eta_c}^2 m^4_{\eta_c} (m_{X^{(\prime)}_2}^2-m_{\eta_c}^2)}{6m_c^2m_{X^{(\prime)}_2}^2}\, ,\nonumber \\
\lambda_{X^{(\prime)}_2J/\psi J/\psi}&=&\frac{\lambda_{X^{(\prime)}_2} f_{J/\psi}^2 m^2_{J/\psi}}{2} \, .
\end{eqnarray}

\section*{Acknowledgements}
This work is supported by National Natural Science Foundation, Grant Number  12175068.

\section*{Authors' contributions}

All authors contributed to the study conception and design. All authors commented on
previous versions of the manuscript. All authors read and approved the final
manuscript.

\section*{Funding}

This research was supported by the National Natural Science Foundation of
China through Grant No.12175068.

\section*{Availability of data and materials}
The data  are available via contacting the corresponding author upon request.

\section*{Competing interests}
The author declare that they have no competing interests.

\end{document}